\documentclass[aps,prl,twocolumn,superscriptaddress,longbibliography]{revtex4-2}
\usepackage{graphicx}
\usepackage{amssymb,amsmath}
\usepackage{bm}
\usepackage{dcolumn}
\usepackage{float}
\usepackage[OT1]{fontenc} 
\usepackage{url}
\usepackage{mathrsfs}
\usepackage{slashed,comment}
\usepackage{color}
\usepackage{verbatim}
\usepackage{enumitem}
\usepackage{soul,physics}
\usepackage[driverfallback=dvipdfm]{hyperref}
\hypersetup{pdfpagemode=FullScreen,colorlinks=true,breaklinks,urlcolor=blue,linkcolor=blue,citecolor=blue}

\usepackage{subfigure}
\usepackage{amssymb}
\usepackage{bm}
\usepackage{graphicx}
\usepackage{amsmath,amssymb}
\usepackage{tikz,fp}
\usepackage{tikz-cd}
\usetikzlibrary{arrows}
\usetikzlibrary{intersections}
\usetikzlibrary{shapes.geometric}
\usetikzlibrary{decorations.pathmorphing, patterns,shapes,fixedpointarithmetic}
\usetikzlibrary{decorations.markings}

\usepackage{pdfpages} % include pdfs
\usepackage{pgffor} % for loops

\makeatletter
\AtBeginDocument{\let\LS@rot\@undefined}
\makeatother

\def\supplementfilename{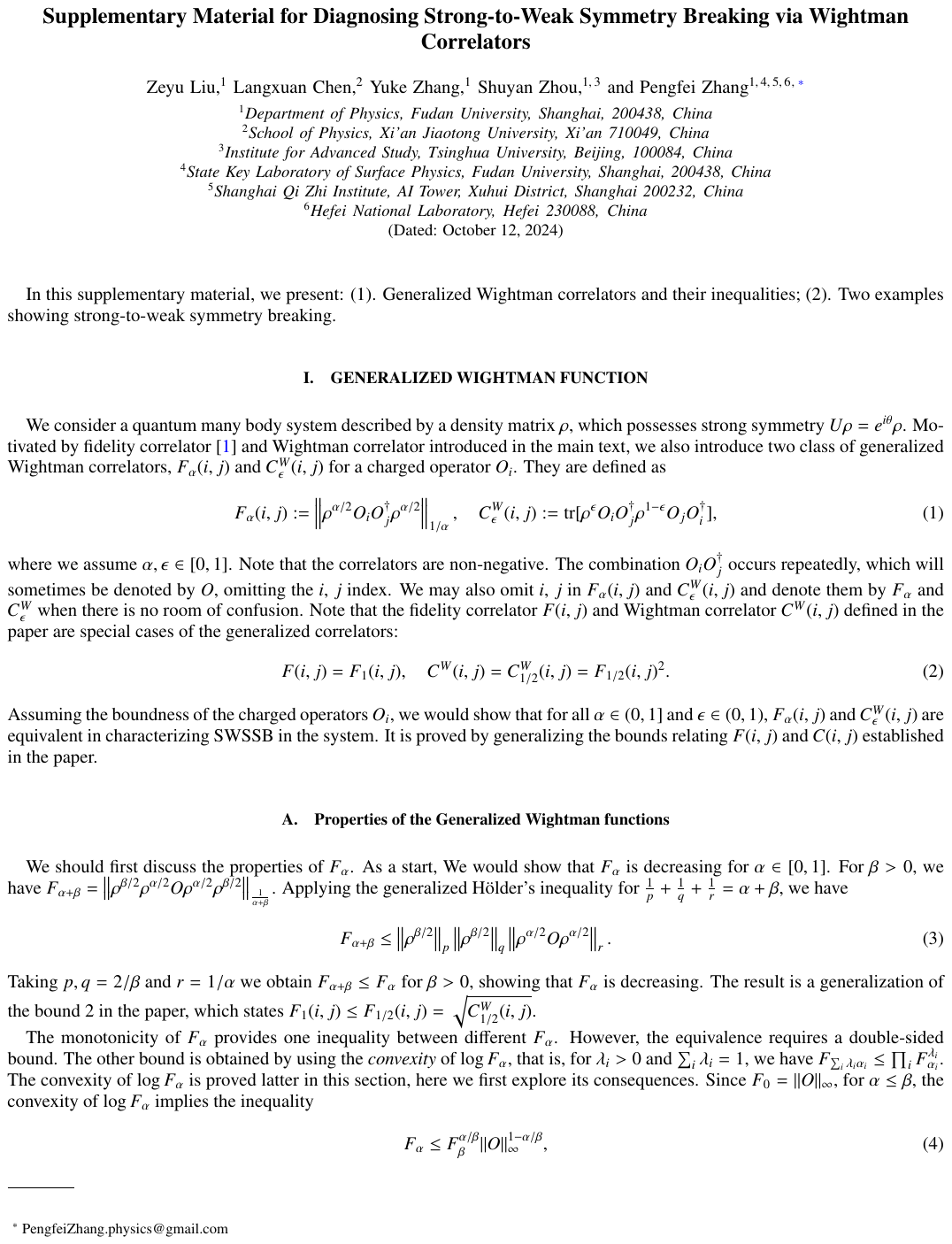}

\pdfximage{\supplementfilename}
\def\numbersupplementpages{\the\pdflastximagepages}

\newif\ifarXiv
\arXivtrue 
\begin{document}
  
  \title{Diagnosing Strong-to-Weak Symmetry Breaking via Wightman Correlators}
  
  \author{Zeyu Liu}
  \affiliation{Department of Physics, Fudan University, Shanghai, 200438, China}

  \author{Langxuan Chen}
  \affiliation{School of Physics, Xi'an Jiaotong University, Xi'an 710049, China}

  \author{Yuke Zhang}
  \affiliation{Department of Physics, Fudan University, Shanghai, 200438, China}

  \author{Shuyan Zhou}
  \affiliation{Department of Physics, Fudan University, Shanghai, 200438, China}
  \affiliation{Institute for Advanced Study, Tsinghua University, Beijing, 100084, China}

  \author{Pengfei Zhang}
  \thanks{PengfeiZhang.physics@gmail.com}
  \affiliation{Department of Physics, Fudan University, Shanghai, 200438, China}
  \affiliation{State Key Laboratory of Surface Physics, Fudan University, Shanghai, 200438, China}
  \affiliation{Shanghai Qi Zhi Institute, AI Tower, Xuhui District, Shanghai 200232, China}
  \affiliation{Hefei National Laboratory, Hefei 230088, China}
  \date{\today}

  \begin{abstract}
  Symmetry plays a fundamental role in quantum many-body physics, and a central concept is spontaneous symmetry breaking, which imposes crucial constraints on the possible quantum phases and their transitions. Recent developments have extended the discussion of symmetry and its breaking to mixed states, enhancing our understanding of novel quantum phases that have no counterpart in pure states. Specific attention has been paid to scenarios where a strongly symmetric density matrix exhibits spontaneous symmetry breaking to weak symmetry, characterized by the fidelity correlator. In this work, we propose the Wightman correlator as an alternative diagnostic tool. This construction relies on the introduction of the thermofield double state for a generic density matrix, which maps the strong symmetry of the density matrix to the doubled symmetry of the pure state, allowing the Wightman correlator to emerge naturally as a standard probe of symmetry breaking. We prove the equivalence between the Wightman function and the fidelity correlator in defining strong-to-weak symmetry breaking, and examine explicit examples involving spin glasses, thermal density matrices, and the decohered Ising model. Additionally, we discuss a susceptibility interpretation of the Wightman correlator.
  \end{abstract}
  
  \maketitle

  \emph{ \color{blue}Introduction.--} Understanding quantum phases and phase transitions is one of the most important topics in quantum many-body physics. A celebrated paradigm is spontaneous symmetry breaking \cite{landau2013statistical}, which posits that different phases are classified based on their symmetry properties, as indicated by the long-range correlation of order parameters $\langle \hat{O}_i\hat{O}^\dagger_j\rangle$. Recent advancements in the study of symmetry and phase transitions in open systems have brought new insights into this fundamental question. An important observation is that the symmetry of density matrices can be defined in two different ways \cite{Buca:2012zz,PhysRevA.89.022118,2018arXiv180200010A,PhysRevLett.125.240405}. As an example, when the particle numbers of the system and the bath are conserved separately, the density matrix $\rho$ can show strong symmetry with $U\rho=e^{i\theta}\rho$, where $U$ generates the symmetry transformation. On the other hand, if the system and bath can exchange particles, the density matrix displays only weak symmetry, fulfilling $U\rho U^\dagger=\rho $. These fruitful symmetry features of open quantum systems promise exciting avenues for exploring novel phases of matter.

  Special attention has been given to the spontaneous breaking of strong symmetry down to weak symmetry, a new symmetry-breaking pattern that has no direct counterpart in pure states \cite{Lessa:2024gcw,Sala:2024ply,Huang:2024rml,Gu:2024wgc,Kuno:2024aiw,Kuno:2024aiw,Zhang:2024fpf,Zhang:2024gbx}. In the seminal work by Lessa et al. \cite{Lessa:2024gcw}, the fidelity correlator is proposed as a universal order parameter for strong-to-weak spontaneous symmetry breaking (SWSSB). It is defined as the fidelity $F(i,j)=\text{tr}[\sqrt{\sqrt{\rho}\sigma\sqrt{\rho}}]$ between the original density matrix $\rho$ and the decorated density matrix $\sigma=\hat{O}_i\hat{O}^\dagger_j \rho \hat{O}_i^\dagger\hat{O}_j$ with charged operator $O_i$. When the fidelity correlator is non-vanishing for $|i-j|\rightarrow \infty$, it is concluded that the system exhibits SWSSB. The SWSSB serves as an obstacle to the disentangling of quantum states with symmetric low-depth quantum channels \cite{Lessa:2024gcw}, and can be understood as a divergence of fidelity susceptibility of local charge dephasing \cite{Zhang:2024gbx}. For systems with strong U(1) symmetry, it is demonstrated that SWSSB implies the hydrodynamic behavior of charges. 

  On the other hand, R\'enyi-2 correlator has been proposed as an alternative candidate for SWSSB, and it has been studied in many examples \cite{Lessa:2024gcw,Sala:2024ply}. One intriguing reason for using the R\'enyi-2 correlator is that its construction is based on the Choi-Jamiolkowski isomorphism \cite{Jamiolkowski:1972pzh,Choi:1975nug} of the density matrix, which maps the density matrix $\rho$ to a pure state $\ket{\rho}\rangle$ in the doubled Hilbert space. Consequently, the symmetry-breaking patterns of the density matrix reduce to those of some well-established pure state. Nevertheless, case studies show that the R\'enyi-2 correlator leads to different transition points compared to the fidelity correlator. Given the series of favorable properties satisfied by the fidelity correlator \cite{Lessa:2024gcw}, the R\'enyi-2 correlator becomes less appealing.

  \begin{figure}[t]
    \centering
    \includegraphics[width=0.85\linewidth]{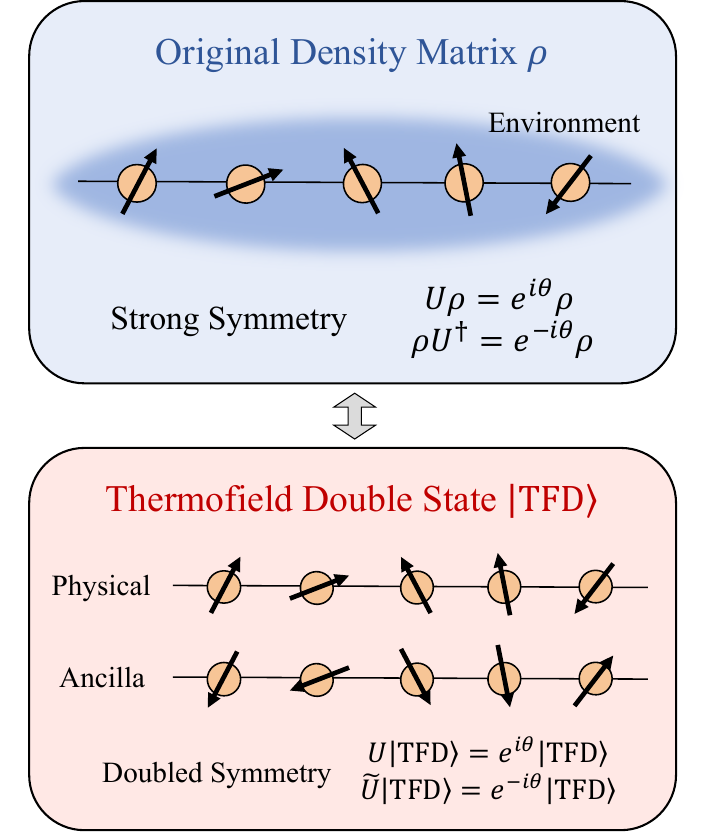}
    \caption{We present a schematic of our proposal for using the Wightman correlator to diagnose strong-to-weak symmetry breaking of the density matrix. The strong symmetry of the density matrix is mapped onto the doubled symmetry of the thermofield double state, with each symmetry acting individually on the physical and auxiliary subsystems. }
    \label{fig:schemticas}
  \end{figure}

  \emph{ \color{blue}Wightman Correlator.--} Our primary aim is to reuse the pure-state perspective of SWSSB. Unlike the Choi-Jamiolkowski isomorphism, which directly employs the operator-state mapping, we consider a specific purification of the density matrix, typically referred to as the thermofield double (TFD) state for thermal density matrices \cite{Israel:1976ur,Maldacena:2001kr}. Here, we extend this terminology to apply to any generic density matrix. We focus on quantum many-body systems that contains $N$ qubits, with Pauli operators $\{X_i,Y_i,Z_i\}$ for $i=1,2,...,N$. The generalization to arbitrary local Hilbert space dimensions and dimensions is straightforward. Now, we introduce $N$ auxiliary qubits with Pauli operators $\{\tilde{X}_i,\tilde{Y}_i,\tilde{Z}_i\}$, and prepare the full system in the (unnormalized) maximally entangled state between the original and the auxiliary system $\ket{\text{EPR}}= \otimes_i\left(\ket{\uparrow\downarrow}_i-\ket{\downarrow\uparrow}_i\right).$ Given any density matrix $\rho$, the corresponding TFD state is defined as
  \begin{equation}
  \ket{\text{TFD}}=(\sqrt{\rho}\otimes I)\ket{\text{EPR}}.
  \end{equation}
  Here, $\sqrt{\rho}$ only operates on the original quantum system. It is straightforward to check that tracing out the auxiliary system reproduce a reduced density matrix $\rho$. 
  
  The symmetry property of the TFD state is inherited from the original density matrix $\rho$. For strongly symmetric density matrices, diagonalization yields $\rho=\sum_a \lambda_a\ket{\psi_a}\bra{\psi_a}$, where all states $\ket{\psi_a}$ possess the same symmetry charge. This leads to
  \begin{equation}
   \ket{\text{TFD}}=\sum_{a}\sqrt{\lambda_a}\ket{\psi_a}\otimes |\tilde{\psi}_a\rangle.
  \end{equation}
  Here, $|\tilde{\psi}_a\rangle:= \bra{\psi_a}\text{EPR}\rangle$ is the state in the auxiliary system.
  To be concrete, let us focus on either the Z$_2$ symmetry with $U=\prod_i X_i$ or U(1) symmetry with $U=\exp(i\phi \sum_i Z_i)$. Using the fact that both $(P_i+\tilde{P}_i)|\text{EPR}\rangle=0$ for arbitrary Pauli operator $P$, we find $U\ket{\psi_a} =e^{i\theta}\ket{\psi_a}$ implies $\tilde{U}|\tilde{\psi}_a\rangle=e^{-i\theta}|\tilde{\psi}_a\rangle$ with unitary transformations on the auxiliary system $\tilde{U}=\prod_i (-1)^N\tilde{X}_i$ or $\tilde{U}=\exp(i\phi \sum_i \tilde{Z}_i)$.
  As a consequence, the physical system and the auxiliary system exhibit doubled symmetry $U$ and $\tilde{U}$. 

  The breaking of symmetry is probed by long-range correlations between charged operators on the TFD state, which can be supported in both physical and auxiliary systems. Let us first consider the scenario where the doubled symmetry is completely broken. Taking the U(1) case as an example, we expect both the charged operator $O_i=S^+_i$ and the auxiliary operator $\tilde{O}_i=\tilde{S}^-_i$ to exhibit long-range correlations. More generally, we have
  \begin{equation}
  \begin{aligned}
  C(i,j)=&\langle O_i O_j^\dagger\rangle_{\text{TFD}}=\langle \tilde{O}_j \tilde{O}_i^\dagger\rangle_{\text{TFD}}=\text{tr}[\rho O_i O_j^\dagger ],
  \end{aligned}
  \end{equation} 
 which corresponds to the standard two-point function of the density matrix $\rho$. Now, if the strong symmetry is broken to weak symmetry, we should choose an operator that carries the charge of $U$ but does not carry the charge of $U\otimes \tilde{U}$. A natural choice is $O_i\tilde{O}_i$, leading to
 \begin{equation}
 C^W(i,j)= \langle O_i\tilde{O}_i O_j^\dagger\tilde{O}_j^\dagger\rangle_{\text{TFD}}.
 \end{equation}
 We expect the SWSSB corresponds to having $C^W(i,j)\neq 0$ for $|i-j|\rightarrow \infty$. This can also be written in the physical Hilbert space as
 \begin{equation}\label{eqn:Wigtman_def}
 C^W(i,j)= \text{tr}\left(\sqrt{\rho}O_iO_j^\dagger\sqrt{\rho}O_i^\dagger O_j\right),
 \end{equation}
 which is non-negative. Again, we adopt the terminology for thermal density matrices and refer to $ C^W(i,j)$ as the \textbf{Wightman correlator}. This correlator has been studied in special cases to probe the SWSSB in \cite{Huang:2024rml}. Compared to the R\'enyi-2 correlator, there are two key differences. Firstly, it replaces $\rho$ in the R\'enyi-2 correlator by $\sqrt{\rho}$. Secondly, no normalization factor is needed since the TFD state is automatically normalized. In the following sections, we prove that the SWSSB defined by the Wightman correlator is equivalent to the definition based on fidelity, and provide illustrative examples from several perspectives. 

 \emph{ \color{blue}Equivalence.--} We first prove that for operators with bounded operator norm $||O||_\infty$, the fidelity correlator is lower bounded by the Wightman correlator with the same operator $O_iO_j^\dagger$. To see this, we first apply the H{\"o}lder's inequality \cite{bhatia97}
 \begin{equation}
 C^W(i,j)\leq ||\sqrt{\rho}O_iO_j^\dagger\sqrt{\rho}||_p\times ||O_i^\dagger O_j||_q,
 \end{equation}
for $1/p+1/q=1$. Here, $||A||_p:= [\text{tr}(A^\dagger A)^\frac{p}{2}]^{1/p}$ denotes the $p$-norm of the operator $A$. We take the limit that $p\rightarrow 1$ and $q\rightarrow \infty$. $||O_i^\dagger O_j||_q$ becomes the $||O||_\infty^2$, while $||\sqrt{\rho}O_iO_j^\dagger\sqrt{\rho}||_1$ exactly matches the fidelity correlator. Therefore, we find the inequality
\begin{equation}\label{eqn:bound1}
 \text{Bound 1:}\ \ \ \ \ \ C^W(i,j)\leq F(i,j)~||O_i||_\infty^2.
 \end{equation}
 In particular, for Pauli operators, $||O_i||_\infty=1$ and we obtain $C^W(i,j)\leq F(i,j)$.
 We proceed to derive how the fidelity correlator serves as a lower bound for the Wightman correlator. To achieve this, we apply the generalized H{\"o}lder's inequality, which states that $||ABC||_1\leq ||A||_p||B||_q||C||_r$ with the constraint $1/p+1/q+1/r=1$. In this context, we set $A=C=\rho^{1/4}$ and $B=\rho^{1/4}O_iO_j^\dagger\rho^{1/4}$. By substituting these definitions into the inequality, we obtain
 \begin{equation}
 F(i,j)\leq ||\rho^{1/4}||_p \times ||\rho^{1/4}||_r \times ||\rho^{1/4}O_iO_j^\dagger\rho^{1/4}||_q.
 \end{equation}
 Let us choose $p=r=4$ and $q=2$. Interestingly, we find $||\rho^{1/4}||_4=(\text{tr}[\rho])^{1/4}=1$, and, by definition, $||\rho^{1/4}O_iO_j^\dagger\rho^{1/4}||_2=\sqrt{C^W(i,j)}$. Therefore, the inequality becomes
 \begin{equation}\label{eqn:bound2}
 \text{Bound 2:}\ \ \ \ \ \ F(i,j)\leq \sqrt{C^W(i,j)}.
 \end{equation}
 By combining both bounds from Eq. \eqref{eqn:bound1} and Eq. \eqref{eqn:bound2}, we conclude that the SWSSB defined by the Wightman correlator is equivalent to the definition using the fidelity correlator. 

 This relation indicates that the Wightman correlator satisfies all the favorable conditions satisfied by the fidelity correlator \cite{Lessa:2024gcw}. For example, the stability condition states that if a density matrix $\rho$ has finite Wightman correlator $C^W(i,j)$ for $|i-j|\rightarrow \infty$ and $\mathcal{E}$ is a strongly symmetric finite-depth local quantum channel, then $\mathcal{E}[\rho]$ also has finite Wightman correlator for $|i-j|\rightarrow \infty$. We also propose several generalizations of our inequality, including two classes of generalized Wightman correlators 
 \begin{equation}
 \begin{aligned}
    &F_{\alpha}(i,j):=\left\lVert \rho^{\alpha/2}O_{i}O_{j}^{\dagger} \rho^{\alpha/2}\right\rVert_{1/\alpha}\ \ \ \ \alpha \in(0,1],
    \\&C^{W}_{\alpha}(i,j):=\text{tr}[\rho^{\alpha} O_{i}O_{j}^{\dagger}\rho^{1-\alpha}O_{i}^{\dagger}O_{j}]\ \ \ \ \alpha \in(0,1).
    \end{aligned}
\end{equation}
 As elaborated in the supplementary material \cite{SM}, both generalizations are equivalent to the fidelity correlator and the Wightman correlator in diagnosing SWSSB. In particular, we have $F_1(i,j)=F(i,j)$ and $F_{1/2}(i,j)=C^W_{1/2}(i,j)=C^W(i,j)$.

 \emph{ \color{blue}Examples.--} Having established generic bounds on the Wightman correlator, we now provide a few examples that illustrate how the Wightman correlator enhances our understanding of SWSSB by simplifying calculations and discovering new connections. More examples can be found in the supplementary material \cite{SM}. 

 \textbf{Spin Glass.} It is known that the SWSSB has a close relationship with traditional discussions of the spin glass \cite{mezard2009information}. We consider special density matrices in which the charged operator cannot couple different states that diagonalize the density matrix $\langle\psi_b|O_i O_j^\dagger|\psi_a\rangle\propto \delta_{ab}$. The fidelity correlator is this scenario becomes $F(i,j)=\sum_a \lambda_a |\langle\psi_a|\hat{O}_i\hat{O}^\dagger_j|\psi_a\rangle|$, which matches the standard definition of the Edwards-Anderson (EA) order parameter \cite{Lessa:2024gcw}. For the Wightman correlator, a straightforward calculation shows that 
 \begin{equation}
 C^W(i,j)=\sum_a \lambda_a |\langle\psi_a|\hat{O}_i\hat{O}^\dagger_j|\psi_a\rangle|^2.
 \end{equation}
 By averaging over $i$ and $j$, we find $N^{-2}\sum_{ij} C^W(i,j):=\chi_{\text{SG}}/N$, where $\chi_{\text{SG}}$ is known as the spin glass susceptibility \cite{mezard2009information}, a universal probe of spin-glass order. It measures the response of a random magnetic field. In the paramagnetic phase, $\chi_{\text{SG}}$ is finite, leading to a vanishing averaged Wightman correlator in the thermodynamic limit. In contrast, in the spin-glass phase, the susceptibility scales with the system size $N$, implying a finite Wightman correlator. Furthermore, higher order moments of $|\langle\psi_a|\hat{O}_i\hat{O}^\dagger_j|\psi_a\rangle|$ can also be probed by generalized Wightman correlators $F_\alpha(i,j)$, while $C^W_\alpha(i,j)$ becomes independent of $\alpha$.

 \textbf{Thermal Ensemble.} It has been conjectured that the finite-temperature canonical ensemble $\rho_{\beta,\text{c}}$ of a local Hamiltonian with charge conservation exhibits SWSSB \cite{Lessa:2024gcw}. Here, we examine the Wightman correlator. Assuming no spontaneous breaking of the weak symmetry, the Wightman function for $|i-j|\gg 1$ can be approximated as
 \begin{equation}
 \begin{aligned}
 &C^W(i,j)= \text{tr}\left(\sqrt{\rho_{\beta,\text{c}}}O_iO_j^\dagger\sqrt{\rho_{\beta,\text{c}}}O_i^\dagger O_j\right)\\
 &\approx\text{tr}\left(\sqrt{\rho_{\beta,\text{gc}}}O_iO_j^\dagger\sqrt{\rho_{\beta,\text{gc}}}O_i^\dagger O_j\right)\approx C_{O_i}^WC_{O_j}^W.
 \end{aligned}
 \end{equation}
 In the first approximation, we replace the canonical ensemble with the grand canonical ensemble $\rho_{\beta,\text{gc}}$, as both yield the same predictions within a fixed charge sector. In the second approximation, we neglect the connected part between operators at site $i$ and $j$. Here, we have introduced the Wightman function of a single charged operator as $C_{O_i}^W=\text{tr}\left(\sqrt{\rho_{\beta,\text{gc}}}O_i\sqrt{\rho_{\beta,\text{gc}}}O_i^\dagger\right)$, which matches the imaginary-time Green's function of $O_i$ with a time separation $\tau=\beta/2$. Since $C_{O_i}^W >0$ for any finite temperature ensemble, we conclude that $\rho_{\beta,\text{c}}$ exhibits the SWSSB. For systems with a charge gap $\Delta$, we expect $C_{O_i}^W \sim e^{-\beta \Delta /2}$, which leads to $C^W(i,j)\sim  e^{-\beta \Delta }$. Interestingly, this matches the result of the fidelity correlator from a replica calculation in \cite{Lessa:2024gcw}. More generally, we have $F_\alpha(i,j)\sim e^{-\beta \Delta }$ and $C_\alpha^W(i,j)\sim e^{-2\tau \Delta }$, with $\tau/\beta=\text{min}\{\alpha ,1-\alpha\}$. Finally, we note that the locality requirement can be relaxed in physical models with all-to-all interactions. A celebrated example is the complex Sachdev-Ye-Kitaev model \cite{Maldacena:2016hyu,Kit.KITP.2,Chowdhury:2021qpy,Davison:2016ngz,Gu:2019jub}, in which the Wightman function has been extensively studied for the calculation of out-of-time-order correlators (OTOC).

 \textbf{Decohered Ising Model.} Another concrete example of SWSSB is the decohered Ising model in 2D \cite{Lessa:2024gcw}, which is dual to the toric code under bit-flip errors \cite{Dennis:2001nw,Wang:2002ph,Chen:2023vxo,Fan:2023rvp,Bao:2023zry}. We consider a 2D square lattice where each site is occupied by a qubit. The system is initialized in the product state along the $x$-direction $\rho_0=\otimes_i\ket{+x}\bra{+x}$. Then, a nearest neighbor Ising channel is applied to the system with 
 \begin{equation}
 \mathcal{E}=\prod_{\langle ij\rangle} \mathcal{E}_{ij},\ \ \ \ \ \ \mathcal{E}_{ij}[\rho_0]=(1-p)+pZ_iZ_j\rho_0 Z_jZ_i.
 \end{equation}
 Here, $p\in[0,1/2]$. The decohered density matrix $\rho=\mathcal{E}[\rho_0]$ exhibits the strong symmetry with $U=\prod_i X_i$. To calculate the Wightman correlator using the replica trick, we first introduce $C^{W,(n)}(i,j)=\text{tr}\left(\rho^{n}Z_iZ_j\rho^{n}Z_i Z_j\right)$. The strategy is to derive results for generic $n$, and perform the analytical continuation to $n=1/2$. Generalizing the analysis in \cite{Lessa:2024gcw}, we find that $C^{W,(n)}(i,j)$ is described by the correlators similar to the fidelity correlator
 \begin{equation}
 \begin{aligned}
 C^{W,(n)}(i,j)=\left\langle \prod_{\alpha=1}^n \sigma_i^{(\alpha)} \sigma_j^{(\alpha)} \right\rangle_{H_{\text{eff}}}.
 \end{aligned}
 \end{equation}
 with
 \begin{equation}
H_{\text{eff}}=-\sum_{\langle ij\rangle}\left(\sum_{\alpha=1}^{2n-1}\sigma_i^{(\alpha)}\sigma_j^{(\alpha)}+\prod_{\alpha=1}^{2n-1}\sigma_i^{(\alpha)}\sigma_j^{(\alpha)}\right),
 \end{equation}
 at an effective temperature $\beta$ determined by $\tanh\beta=p/(1-p)$. In the single replica limit $n\rightarrow 1/2$, the effective model describes the random bond Ising model along the Nishimori line \cite{Fan:2023rvp}, which undergoes a ferromagnetic-to-paramagnetic phase transition. The SWSSB for $\rho$ occurs when the random bond Ising model becomes ordered at $p>p_c\approx 0.109$. Here, we emphasize the significance of the single replica limit, enabled by the use of the TFD state, or equivalently, by having $\sqrt{\rho}$ in Eq. \eqref{eqn:Wigtman_def}. This also implies that the same transition point applies to all generalized Wightman correlators.

 \emph{ \color{blue}Susceptibility Interpretation.--} Motivated by its close relation to spin glass susceptibility, we present an interpretation of the Wightman correlator in terms of susceptibility in generic setups. There is a similar proposal for the fidelity correlator \cite{Zhang:2024gbx}. We consider the Wightman function $C_{O_i}^W=\langle O_i\tilde{O}_i^\dagger\rangle_{\text{TFD}}$ of a single charged operator $O_i$ for a generic density matrix. When $\rho$ exhibits strong symmetry, the Wightman function vanishes due to the charge constraint. Now, we introduce a perturbation to the TFD state by coupling the physical system and the auxiliary system with a direct hopping
 \begin{equation}
 |\psi\rangle\sim \exp\Big(\frac{\epsilon}{2}\sum_j \left[O_j^\dagger\tilde{O}_j+O_j\tilde{O}_j^\dagger\right]\Big)\ket{\text{TFD}}.
 \end{equation}
 Physically, this perturbation enables charge fluctuations. We then measure the change of the Wightman function $C_{O_i}^W$. Perturbatively, the result is given by
 \begin{equation}
 C_{O_i}^W/ \epsilon:=\chi^W= \sum_j C^W(i,j).
 \end{equation}
 Therefore, the susceptibility $\chi^W$ remains finite when the strong symmetry is unbroken but diverges if SWSSB occurs. This is closely analogous to the spin glass case. To leading order in $\epsilon$, we can translate the perturbation into the language of quantum channel $\mathcal{N}=\prod_i\mathcal{N}_i$ with:
 \begin{equation}
 \mathcal{N}_i[\rho]=\left(1-\frac{\epsilon^2}{2}\right)\rho+\frac{\epsilon^2}{4}O_i^\dagger \rho O_i+\frac{\epsilon^2}{4}O_i \rho O_i^\dagger.
 \end{equation}
 Here, we assume the charged operator $O_i$ is unitary. This quantum channel perturbs the density matrix by a homogeneous charge creation.

  We can also interpret the generalized Wightman functions as susceptibilities using a similar construction. Here, we briefly discuss an alternative strategy for $C_\alpha^W(i,j)$ with $\alpha \rightarrow 0$, showing a close relation to the response of von Neumann entropy. Assuming that $O_i$ is unitary, we have
 \begin{equation}
 C_\alpha^W(i,j)=C_0^W(i,j)-\alpha~\text{tr}[H_M\delta \rho],
 \end{equation}
 to the leading order in $\alpha$. Here, we introduce the modular Hamiltonian $H_M=-\ln \rho$ and $\delta \rho =  [O_{i}O_{j}^{\dagger}\rho O_{i}^{\dagger}O_{j}-\rho]$. The second term can be interpreted as computing the difference in von Neumann entropy between the original density matrix $\rho$ and the decohered density matrix 
 \begin{equation}
     \mathcal{M}[\rho]=(1-\alpha)\rho+ \alpha O_{i}O_{j}^{\dagger}\rho O_{i}^{\dagger}O_{j}=\rho+\alpha \delta\rho,
 \end{equation}
 in the limit of $\alpha \rightarrow 0$. For systems without SWSSB, $C_0^W(i,j)>0$ while $C_\alpha^W(i,j)=0$ for $\alpha>0$, indicating a divergence in the entropy response $\text{tr}[H_M\delta \rho]$. This further suggests that the modular Hamiltonian of density matrices without SWSSB should exhibit spooky properties, possibly with operator sizes scaling with the system size or an unbounded spectrum. This relation also provides solid proof that the thermal state of physical Hamiltonians always exhibits SWSSB, for which the entropy response function is bounded.

 \emph{ \color{blue}Discussions.--} In this work, we introduce the use of the Wightman function to diagnose SWSSB. Our approach involves mapping the input density matrix to the thermofield double state, where the strong symmetry is transformed into a doubled symmetry, represented by $U$ and $\tilde{U}$. The Wightman function naturally arises as a correlator of charged operators within this framework. By establishing two-sided bounds, we demonstrate the equivalence between the Wightman correlator and the fidelity correlator, with several illustrative examples provided. We also draw an analogy with spin glass susceptibility to propose a susceptibility interpretation of the Wightman correlator.

 The validity of this pure-state perspective has significant implications for understanding SWSSB. It is well-known that order-to-disorder transitions can be probed by the expectation value of disorder operators, which remain non-zero in the disordered phase \cite{Fradkin:2016ksx}. We anticipate that a similar disorder operator can be introduced straightforwardly within our TFD setup. Furthermore, recent studies highlight the singularity of the full counting statistics of charges in a subregion during the superfluid phase \cite{Wang:2023csw}. Applied to the TFD state, this framework suggests the singularity of a correlator involving charge twist operators acting on both the physical and auxiliary systems.
 Another intriguing direction is exploring SWSSB from the perspective of the modular flow. For system with SWSSB, our analysis of entropy response suggests the modular Hamiltonian is well-behaved. This facilitates the study of modular flow, which defines the intrinsic dynamics of the density matrix. For instance, evolving the operator $O_j$ in Eq. \eqref{eqn:Wigtman_def} to a later time $t$ results in an OTOC. This framework could pave the way for further classification of density matrices, allowing for the distinction between those with chaotic and non-chaotic modular Hamiltonians.

 \vspace{5pt}
  \textit{Acknowledgments.}
  We thank Ning Sun, Zijian Wang, Zhaoyi Zeng, and Tian-Gang Zhou for helpful discussions. We are especially grateful to Jian-Hao Zhang for bringing this topic to our attention and for offering invaluable explanations and discussions. This project is supported by the NSFC under grant number 12374477 and by Innovation Program for Quantum Science and Technology under grant number 2024ZD0300100.

\bibliography{Wightmann_Correlator.bbl}

\ifarXiv


\section*{Supplementary material}

# Supplementary Material for Diagnosing Strong-to-Weak Symmetry Breaking via Wightman Correlators

Zeyu Liu,[1] Langxuan Chen,[2] Yuke Zhang,[1] Shuyan Zhou,[1,3] and Pengfei Zhang[1,4,5,6,∗]

[1]*Department of Physics, Fudan University, Shanghai, 200438, China*
[2]*School of Physics, Xi'an Jiaotong University, Xi'an 710049, China*
[3]*Institute for Advanced Study, Tsinghua University, Beijing, 100084, China*
[4]*State Key Laboratory of Surface Physics, Fudan University, Shanghai, 200438, China*
[5]*Shanghai Qi Zhi Institute, AI Tower, Xuhui District, Shanghai 200232, China*
[6]*Hefei National Laboratory, Hefei 230088, China*
(Dated: October 12, 2024)

In this supplementary material, we present: (1). Generalized Wightman correlators and their inequalities; (2). Two examples showing strong-to-weak symmetry breaking.

## I. GENERALIZED WIGHTMAN FUNCTION

We consider a quantum many body system described by a density matrix $\rho$, which possesses strong symmetry $U\rho = e^{i\theta}\rho$. Motivated by fidelity correlator [1] and Wightman correlator introduced in the main text, we also introduce two class of generalized Wightman correlators, $F_\alpha(i, j)$ and $C_\epsilon^W(i, j)$ for a charged operator $O_i$. They are defined as

$$F_\alpha(i, j) := \left\| \rho^{\alpha/2} O_i O_j^\dagger \rho^{\alpha/2} \right\|_{1/\alpha}, \quad C_\epsilon^W(i, j) := \text{tr}[\rho^\epsilon O_i O_j^\dagger \rho^{1-\epsilon} O_j O_i^\dagger], \tag{1}$$

where we assume $\alpha, \epsilon \in [0, 1]$. Note that the correlators are non-negative. The combination $O_i O_j^\dagger$ occurs repeatedly, which will sometimes be denoted by $O$, omitting the $i, j$ index. We may also omit $i, j$ in $F_\alpha(i, j)$ and $C_\epsilon^W(i, j)$ and denote them by $F_\alpha$ and $C_\epsilon^W$ when there is no room of confusion. Note that the fidelity correlator $F(i, j)$ and Wightman correlator $C^W(i, j)$ defined in the paper are special cases of the generalized correlators:

$$F(i, j) = F_1(i, j), \quad C^W(i, j) = C_{1/2}^W(i, j) = F_{1/2}(i, j)^2. \tag{2}$$

Assuming the boundness of the charged operators $O_i$, we would show that for all $\alpha \in (0, 1]$ and $\epsilon \in (0, 1)$, $F_\alpha(i, j)$ and $C_\epsilon^W(i, j)$ are equivalent in characterizing SWSSB in the system. It is proved by generalizing the bounds relating $F(i, j)$ and $C(i, j)$ established in the paper.

### A. Properties of the Generalized Wightman functions

We should first discuss the properties of $F_\alpha$. As a start, We would show that $F_\alpha$ is decreasing for $\alpha \in [0, 1]$. For $\beta > 0$, we have $F_{\alpha+\beta} = \left\| \rho^{\beta/2} \rho^{\alpha/2} O \rho^{\alpha/2} \rho^{\beta/2} \right\|_{\frac{1}{\alpha+\beta}}$. Applying the generalized Hölder's inequality for $\frac{1}{p} + \frac{1}{q} + \frac{1}{r} = \alpha + \beta$, we have

$$F_{\alpha+\beta} \leq \left\| \rho^{\beta/2} \right\|_p \left\| \rho^{\beta/2} \right\|_q \left\| \rho^{\alpha/2} O \rho^{\alpha/2} \right\|_r. \tag{3}$$

Taking $p, q = 2/\beta$ and $r = 1/\alpha$ we obtain $F_{\alpha+\beta} \leq F_\alpha$ for $\beta > 0$, showing that $F_\alpha$ is decreasing. The result is a generalization of the bound 2 in the paper, which states $F_1(i, j) \leq F_{1/2}(i, j) = \sqrt{C_{1/2}^W(i, j)}$.

The monotonicity of $F_\alpha$ provides one inequality between different $F_\alpha$. However, the equivalence requires a double-sided bound. The other bound is obtained by using the *convexity* of $\log F_\alpha$, that is, for $\lambda_i > 0$ and $\sum_i \lambda_i = 1$, we have $F_{\sum_i \lambda_i \alpha_i} \leq \prod_i F_{\alpha_i}^{\lambda_i}$. The convexity of $\log F_\alpha$ is proved latter in this section, here we first explore its consequences. Since $F_0 = \|O\|_\infty$, for $\alpha \leq \beta$, the convexity of $\log F_\alpha$ implies the inequality

$$F_\alpha \leq F_\beta^{\alpha/\beta} \|O\|_\infty^{1-\alpha/\beta}, \tag{4}$$

---

∗ PengfeiZhang.physics@gmail.com

which generalizes bound 1 in the paper. Taking $\alpha = 1/2$ and $\beta = 1$, we recover the bound $C^W(i, j) \leq F(i, j) \|O_i\|_\infty^2$.

In summary, for $F_\alpha$ we have the inequality

$$F_\beta \leq F_\alpha \leq F_\beta^{\alpha/\beta} \|O\|_\infty^{1-\alpha/\beta}, \tag{5}$$

which shows that for any pair of $\alpha, \beta \in (0, 1]$, $\lim_{|i-j|\to\infty} F_\alpha(i, j) = 0$ if and only if $\lim_{|i-j|\to\infty} F_\beta(i, j) = 0$. As a result, $F_\alpha$ and $F_\beta$ are equivalent in characterizing SWSSB.

Having finished the discussion of $F_\alpha(i, j)$, we proceed to discuss the correlators $C_\epsilon^W(i, j)$ for $\epsilon \in (0, 1)$. We would establish the equivalence between $C_\epsilon^W(i, j)$ and some $F_\alpha(i, j)$, hence showing that all Generalized Wightman functions (for $0 < \alpha \leq 1$ and $0 < \epsilon < 1$) are equivalent in characterizing SWSSB. For $\epsilon \in (0, 1/2]$, An upper bound for $C_\epsilon^W(i, j)$ is provided as

$$C_\epsilon^W = \text{tr}[\rho^\epsilon O \rho^\epsilon \rho^{1-2\epsilon} O^\dagger] \leq \|\rho^\epsilon O \rho^\epsilon\|_p \|\rho^{1-2\epsilon}\|_q \|O\|_\infty, \tag{6}$$

where $p, q > 0$ and $1/p + 1/q = 1$. Choosing $p = 1/2\epsilon$ and $q = 1/(1 - 2\epsilon)$, we obtain the upper bound. $C_\epsilon^W \leq F_{2\epsilon} \|O\|_\infty$. Similarly, for $\epsilon \in [1/2, 1)$ we have $C_\epsilon^W \leq F_{2(1-\epsilon)} \|O\|_\infty$.

To obtain a lower bound, note that

$$C_\epsilon^W = \text{tr}[\rho^{\frac{\epsilon}{2}} O \rho^{\frac{1-\epsilon}{2}} \rho^{\frac{1-\epsilon}{2}} O^\dagger \rho^{\frac{\epsilon}{2}}] = \|\rho^{\frac{\epsilon}{2}} O \rho^{\frac{1-\epsilon}{2}}\|_2^2. \tag{7}$$

Taking the special case $C_{1/2}^W$:

$$F_{1/2}^2 = C_{1/2}^W = \text{tr}[\rho^{\frac{\epsilon}{2}} O \rho^{\frac{1-\epsilon}{2}} \rho^{\frac{\epsilon}{2}} O^\dagger \rho^{\frac{1-\epsilon}{2}}] \leq \|\rho^{\frac{\epsilon}{2}} O \rho^{\frac{1-\epsilon}{2}}\|_2 \|\rho^{\frac{1-\epsilon}{2}} O \rho^{\frac{\epsilon}{2}}\|_2 = \sqrt{C_\epsilon^W C_{1-\epsilon}^W}. \tag{8}$$

The lower bound for $C_\epsilon^W$ is then obtained by noticing that $C_{1-\epsilon}^W$ is bounded from above by some $F_\alpha$, which concludes the proof that the Generalized Wightman functions are equivalent for characterizing SWSSB. However, the detailed lower bound for $C_\epsilon$ is rather complicated. It is more convenient to work with the geometric mean $\tilde{C}_\epsilon^W = \sqrt{C_\epsilon^W C_{1-\epsilon}^W}$, for which we have the inequalities

$$F_{1/2}^2 \leq \tilde{C}_\epsilon \leq F_{2\epsilon} \|O\|_\infty, \tag{9}$$

where we assume $\epsilon \in (0, 1/2]$. We may choose to bound both side by the original fidelity correlator $F_1$, which leads to the simple double bound

$$F(i, j)^2 \leq \tilde{C}_\epsilon(i, j) \leq F(i, j)^{2\epsilon} \|O\|_\infty^{2(1-\epsilon)}. \tag{10}$$

## B. Proving the convexity of the function $\log F_\alpha$

In this section we prove that $\log F_\alpha$ is convex for $\alpha \in [0, 1]$. The inequality we need is a slight variation of the Hölder's inequality, but for matrices there are many similar variations of the Hölder's inequality and it's not easy to find a good reference. As a result, we shall provide a self-contained proof.

We need to introduce some necessary tools and conventions before start. In what follows, the norm symbol $\|\cdot\|$ is sometimes used without specifying $p$, in which case it is always understood to be $\|\cdot\|_\infty$. The convention is made mainly because $\|\cdot\|_\infty$ is the most common operator norm.

Suppose $A$ is a linear operator on a Hilbert Space $\mathcal{H}$. Its *singular values* $s_i(A)$ are defined as eigenvalues of $\sqrt{A^\dagger A}$, ordered so that $s_1(A) \geq s_2(A) \geq \ldots \geq s_n(A)$. Its *eigenvalues* $\lambda_i(A)$ are defined as usual, and are also ordered as $|\lambda_1(A)| \geq |\lambda_2(A)| \geq, \ldots \geq |\lambda_n(A)|$. Note that even for diagonalizable operator $A$, $s_i(A) \neq |\lambda_i(A)|$ in general if $A$ is not Hermitian. The operator $p$ norm of $A$ is defined as $\|A\|_p = (\sum_i s_i(A)^p)^{1/p}$ for $p \geq 1$. The definition might be extended to all $p > 0$, but $\|\cdot\|_p$ is not a norm for $p < 1$. There are important inequalities between $\lambda_i(A)$ and $s_i(A)$, as we should explore now. Consider the subspace of all completely anti-symmetric tensors in $\otimes_{i=1}^k \mathcal{H}$, which is usually denoted $\wedge^k \mathcal{H}$. Being a subspace of $\otimes_{i=1}^k \mathcal{H}$, $\wedge^k \mathcal{H}$ is naturally a Hilbert space. Now consider the operator $\otimes_{i=1}^k A$ acting on $\otimes_{i=1}^k \mathcal{H}$, it's clear that it maps $\wedge^k \mathcal{H}$ to itself. Define the corresponding action of $\otimes_{i=1}^k A$ on $\wedge^k \mathcal{H}$ as $\wedge^k A$. The ($p = \infty$) norm of $\wedge^k A$ satisfies

$$\|\wedge^k A\| = \prod_{i=1}^k s_i(A). \tag{11}$$

This is true because the $p$ norm is invariant under multiplications of unitary operators. One may replace $A$ by $UA$ for unitary operator $U$ on $\mathcal{H}$, such that $UA$ is positive definite with eigenvalues $s_i(A)$. The eigenvalues of the positively definite operator

$\wedge^k UA = \wedge^k U \cdot \wedge^k A$ would be given by $s_{i_1}(A)s_{i_2}(A)\dots s_{i_k}(A)$, where $\{i_1, \dots, i_k\}$ are distinct values in $\{1, \dots, n\}$. Note that $\wedge^k U$ is unitary on $\wedge^k \mathcal{H}$ since $\otimes_{i=1}^k U$ is unitary on $\otimes_{i=1}^k \mathcal{H}$. As a result we see $\|\wedge^k A\| = \|\wedge^k UA\| = \prod_{i=1}^k s_i(A)$. On the other hand, the operator $\wedge^k A$ has eigenvalues $\lambda_{i_1}(A)\lambda_{i_2}(A)\dots\lambda_{i_k}(A)$ for distinct $i_1, \dots, i_k$, which leads to the inequality

$$\prod_{i=1}^k |\lambda_i(A)| \le \|\wedge^k A\| = \prod_{i=1}^k s_i(A), \quad 1 \le k \le \dim A. \tag{12}$$

For operators $A$ and $B$, we have $\wedge^k A \cdot \wedge^k B = \wedge^k AB$ because $\otimes_{i=1}^k A \cdot \otimes_{i=1}^k B = \otimes_{i=1}^k AB$. As a result, we have the bound

$$\prod_{i=1}^k s_i(AB) = \|\wedge^k AB\| \le \|\wedge^k A\| \times \|\wedge^k B\| = \prod_{i=1}^k s_i(A)s_i(B), \quad 1 \le k \le \dim A, B. \tag{13}$$

To prove the convexity we need yet another technique called *Marjorization*. Suppose $\{a_i\}_{i=1}^n$, $\{b_i\}_{i=1}^n$ are sequences of real numbers such that $a_1 \ge a_2 \ge \dots \ge a_n$, $b_1 \ge b_2 \ge \dots \ge b_n$ and for $1 \le k \le n$, $\sum_{i=1}^k a_i \ge \sum_{i=1}^k b_i$, then we say the sequence $a_i$ *Majorizes* $b_i$, commonly denoted $a \succ b$. It's then not hard to prove that for $r > 0$ we have $\sum_i e^{ra_i} \ge \sum_i e^{rb_i}$ by using the convexity and monotonicity of the exponential function $e^x$. With the above preparation, Hölder's inequality for operators might be proved immediately. From Eq. (13), we see that $a_i = \log s_i(A)s_i(B)$ majorizes $b_i = \log s_i(AB)$, thus for any $r > 0$,

$$\sum_i s_i(AB)^r = \sum_i \exp^{r\log s_i(AB)} \le \sum_i \exp^{r\log s_i(A)s_i(B)} = \sum_i s_i(A)^r s_i(B)^r. \tag{14}$$

The Hölder's inequality for operators then follows from the ordinary Hölder's inequality for functions. Proving the convexity of $F_\alpha$ is very similar. What we would show is that for $\alpha, \beta \in [0,1]$, the following inequality hold:

$$F_{\frac{\alpha+\beta}{2}}^2 \le F_\alpha F_\beta. \tag{15}$$

It guarantees the convexity of $\log F_\alpha$ since it leads to the inequality $\log F_{\frac{\alpha+\beta}{2}} \le \frac{1}{2}(\log F_\alpha + \log F_\beta)$. To prove the above relation, note that

$$F_{\frac{\alpha+\beta}{2}}^2 = \left\| \rho^{\frac{\alpha+\beta}{4}} O \rho^{\frac{\alpha+\beta}{4}} \right\|_{\frac{2}{\alpha+\beta}}^2 = \mathrm{tr}\left[ \left( \sqrt{\rho^{\frac{\alpha+\beta}{4}} O \rho^{\frac{\alpha+\beta}{2}} O^\dagger \rho^{\frac{\alpha+\beta}{4}}} \right)^{\frac{2}{\alpha+\beta}} \right]^{\alpha+\beta} = \mathrm{tr}\left[ \left( \sqrt{\rho^{\frac{\alpha}{2}} O \rho^{\frac{\alpha}{2}} \rho^{\frac{\beta}{2}} O^\dagger \rho^{\frac{\beta}{2}}} \right)^{\frac{2}{\alpha+\beta}} \right]^{\alpha+\beta}, \tag{16}$$

where the cyclic property of the trace is used. Setting $r = \frac{2}{\alpha+\beta}$, $A = \rho^{\frac{\alpha}{2}} O \rho^{\frac{\alpha}{2}}$, $B = \rho^{\frac{\beta}{2}} O^\dagger \rho^{\frac{\beta}{2}}$, we have $F_{\frac{\alpha+\beta}{2}}^2 = \mathrm{tr}\left[(AB)^{r/2}\right]^{2/r}$, where $r \ge 1$ and $AB$ is diagonalizable with eigenvalues $\lambda_i(AB) \ge 0$ for all $i$. Focusing on the operator $\sqrt{AB}$. For $1 \le k \le \dim\sqrt{AB}$,

$$\prod_{i=1}^k \lambda_i\left(\sqrt{AB}\right) = \prod_{i=1}^k \sqrt{\lambda_i(AB)} \le \prod_{i=1}^k \sqrt{s_i(AB)} \le \prod_{i=1}^k \sqrt{s_i(A)}\sqrt{s_i(B)}. \tag{17}$$

Similar to our proof of the Hölder's inequality for operators,

$$\mathrm{tr}\left[(AB)^{r/2}\right] = \sum_i \lambda_i\left(\sqrt{AB}\right)^r \le \sum_i \left(\sqrt{s_i(A)}\sqrt{s_i(B)}\right)^r \le \|A\|_{p/2}^{r/2} \|B\|_{q/2}^{r/2}, \tag{18}$$

where $p, q$ are positive constants with $1/p + 1/q = 1/r$. Since $r = \frac{2}{\alpha+\beta}$, we may choose $p = \frac{2}{\alpha}$, $q = \frac{2}{\beta}$. Combine Eq. (16) and Eq. (18) we have

$$F_{\frac{\alpha+\beta}{2}}^2 = \mathrm{tr}\left[(AB)^{r/2}\right]^{2/r} \le \|A\|_{p/2}\|B\|_{q/2} = \left\|\rho^{\alpha/2} O \rho^{\alpha/2}\right\|_{1/\alpha} \left\|\rho^{\beta/2} O^\dagger \rho^{\beta/2}\right\|_{1/\beta} = F_\alpha F_\beta, \tag{19}$$

which proves that $\log F_\alpha$ is a convex function.

## II. MORE ILLUSTRATIVE EXAMPLES

Recall that a mixed state with strong symmetry exhibits strong-to-weak spontaneous symmetry breaking (SWSSB) in the sense of the Wightman correlator if

$$C^W(i,j) = \mathrm{tr}\left(\sqrt{\rho} O_i O_j^\dagger \sqrt{\rho} O_i^\dagger O_j\right) \ne 0, \tag{20}$$

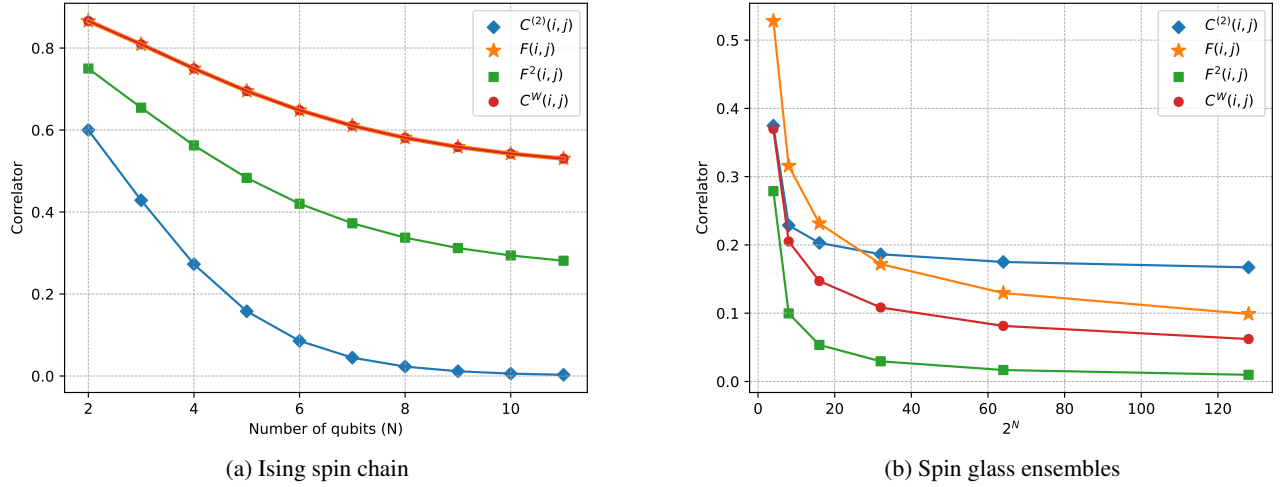

(a) Ising spin chain

(b) Spin glass ensembles

FIG. 1. Rényi-2 correlator $C^{(2)}(i, j)$, fidelity correlator $F(i, j)$, square of fidelity correlator $F^2(i, j)$ and Wightman correlator $C^W(i, j)$ as a function of number of qubits for (a). Ising spin chain with density matrix (24) and (b). Spin glass ensembles with density matrix (27).

for some charged local operator $O_i$ as $|i - j| \to \infty$. At the same time, for any charged local operator $O'_i$,

$$\text{tr}\left(\rho O'_i O'^\dagger_j\right) \to 0, \tag{21}$$

as $|i - j| \to \infty$. Other diagnostics that can identify strong-to-weak spontaneous symmetry breaking (SWSSB) include the fidelity correlator [1]

$$F(i, j) = \text{tr}\left(\sqrt{\sqrt{\rho} O_i O^\dagger_j \rho O^\dagger_i O_j \sqrt{\rho}}\right). \tag{22}$$

Analogous to Eq. (20), one can define SWSSB based on fidelity correlator: a state $\rho$ has SWSSB when the fidelity correlator is non-vanishing while the conventional correlation function vanishes. In the main text and Section I, we explain that Wightman correlator and the fidelity correlator are equivalent in their characterization of SWSSB. Additionally, it is important to note that there is also an alternative definition, based on the Rényi-2 correlator [1, 2]

$$C^{(2)}(i, j) = \frac{\text{tr}\left(\rho O_i O^\dagger_j \rho O^\dagger_i O_j\right)}{\text{tr}\left(\rho^2\right)}, \tag{23}$$

which does not possess this equivalence.

To illustrate the differences between these correlators, we will consider two examples propsed in [1]. First, let us consider an Ising spin chain with density matrix

$$\rho = \frac{1}{2} \otimes_i |+x\rangle \langle +x| + \frac{1}{2^{N+1}} (I + U), \tag{24}$$

where $N$ is the number of qubits and $U = \prod_i X_i$ is the unitary operator that generates the symmetry transformation. It is straightforward to check that this state has strong symmetry since $U\rho = \rho$. For charged operator $Z_i$, we have

$$F(i, j) = C^W(i, j) = \frac{1}{2} + \frac{1}{2^{N-1}}\left(\left(2^{N-1} + 1\right)^{1/2} - 1\right), \quad C^{(2)}(i, j) = \frac{3}{2^{N-1} + 3}, \tag{25}$$

as long as $i \neq j$. As $N \to \infty$, we have $F(i, j) = C^W(i, j) \to 1/2$ and $C^{(2)}(i, j) \to 0$, indicating that the state $\rho$ exhibits SWSSB in the sense of the Wightman (fidelity) correlator, but not in the sense of the Rényi-2 correlator. The numerical results are shown in FIG. 1 (a). Conversely, a situation where $C^W(i, j) \to 0$ but $C^{(2)}(i, j) \neq 0$ can also occur, indicating SWSSB only in the sense of the Rényi-2 correlator. Before delving into the concrete example, let us first examine the following bound:

$$\text{tr}\left(\rho O_i O^\dagger_j \rho O^\dagger_i O_j\right) = \left\|\rho^{1/2} O_i O^\dagger_j \rho^{1/2}\right\|_2^2 \leq \left\|\rho^{1/2} O_i O^\dagger_j \rho^{1/2}\right\|_1^2 = F^2(i, j) \leq C^W(i, j) \leq F(i, j), \tag{26}$$

when $O_i$ is unitary. The first equality follows directly from the definition of the 2-norm given in Section I. The second inequality holds because the $p$-norm is monotonically decreasing. The remaining relations follow from the bound in the main text for completeness. Therefore, the numerator in (23) is less than the Wightman function. Therefore, we conclude that a necessary condition for the system to have SWSSB only in the sense of Rényi 2 correlator is $\mathrm{tr}\left(\rho^2\right) \to 0$, which makes it less appealing. One such example is provided below.

Consider a qubit chain with $2N$ spins, focusing on a $Z_2$ symmetry that only acts on the even sites, generated by $U_e = \otimes_n X_{2n}$. Next, we consider all length-$N$ bit strings $m_1 \cdots m_N$, where each $m_i$ equals either 0 or 1, and define $m = \sum_{i=1}^{N} 2^{N-i} m_i$. We then introduce the unnormalized density matrix as

$$\rho = \sum_{m=1}^{2^N} p_m |\psi_m\rangle\langle\psi_m|, \quad p_m = m^{-2/3},$$ (27)

where

$$|\psi_m\rangle = \frac{I + U_e}{2} |m_1 \tilde{m}_1 m_2 \tilde{m}_2 \cdots\rangle, \quad |\tilde{m}_i\rangle = \sqrt{\frac{1}{2} + \frac{1}{m}} |m_i\rangle + \sqrt{\frac{1}{2} - \frac{1}{m}} X_i |m_i\rangle.$$ (28)

The state $|m_i\rangle$, defined on the odd sites, ensures that different $|\psi_m\rangle$ are orthogonal. For large m, we have

$$|\tilde{m}_i\rangle \approx |+x\rangle + [(-1)^{m_i}/m] |-x\rangle, \quad Z_i|\tilde{m}_i\rangle \approx |-x\rangle + [(-1)^{m_i}/m] |+x\rangle.$$ (29)

Thus, we obtain

$$\langle\psi_m|Z_i Z_j|\psi_n\rangle = z_m \delta_{m,n} \sim \frac{1}{m^2}\delta_{m,n}.$$ (30)

for $i \neq j$, where $Z_i$ is only defined on the even sites. Then it becomes straightforward to show that

$$C^{(2)}(i,j) = \frac{\sum_m p_m^2 |z_m|^2}{\sum_m p_m^2}, \quad F(i,j) = \frac{\sum_m p_m |z_m|}{\sum_m p_m}, \quad C^{(W)}(i,j) = \frac{\sum_m p_m |z_m|^2}{\sum_m p_m}.$$ (31)

Use the fact that $p_m \sim m^{-2/3}$ and $z_m \sim m^{-2}$, we find that $\sum_m p_m^2 |z_m|^2$, $\sum_m p_m^2$, $\sum_m p_m |z_m|$ and $\sum_m p_m |z_m|^2$ all converge to a finite value, while $\sum_m p_m$ diverges. Thus, we expect that the Wightman (fidelity) correlator is exponentially small, while the Rényi-2 correlator is finite. The numerical results are shown in FIG. 1 (b).

---


\fi

\end{document}